\documentclass[a4paper, 12pt]{article}
\usepackage{graphics, amsmath, amssymb, array, cancel, cite}

\newcommand{\me}{\mathrm{e}}
\newcommand{\mi}{\mathrm{i}}
\newcommand{\mop}[1]{\operatorname{#1}}
\newcommand{\dif}{\mathrm{d}}
\newcommand{\sdif}[1]{\!\!\dif#1\;}
\newcommand{\simm}[1]{\mathrel{\vcenter{\hbox{$#1$}\nointerlineskip\hbox{$\sim$}}}}

\begin{document}

\begin{titlepage}

\begin{flushright}
  IPPP/03/69\\
  DCPT/03/138\\
  hep-th/0311051\\
  November 2003
\end{flushright}

\begin{center}
  \vspace{15ex}
  {\bf\Large Brane-Antibrane Kinetic Mixing, Millicharged Particles and SUSY Breaking\\}
  \vspace{6ex}
  {\bf S. A. Abel and B. W. Schofield\\}
  \vspace{6ex}
  {\it Institute for Particle Physics Phenomenology and Department of Mathematical Sciences\\
       University of Durham, Durham, DH1 3LE, UK\\}
  \vspace{2ex}
  {\tt s.a.abel@durham.ac.uk, b.w.schofield@durham.ac.uk\\}
  \vspace{15ex}
  
  \parbox{140mm}{{\bf\centering Abstract\\} \vspace{2ex} It is known that hidden $U(1)$ gauge factors can
    couple to visible $U(1)$'s through Kinetic Mixing. This phenomenon is shown generically to occur in
    nonsupersymmetric string set-ups, between D-branes and $\overline{\text{D}}$-branes. Kinetic Mixing
    acts either to give millicharges (of e.g. hypercharge) to would-be hidden sector fermions, or to
    generate an enhanced communication of supersymmetry breaking that dominates over the usual
    gravitational suppression.  In either case, the conclusion is that the string scale in
    nonsupersymmetric brane configurations has a generic {\em upper} bound of $M_s \simm{<} 10^{8}
    \,\text{GeV}$.}
\end{center}

\end{titlepage}


\section{Introduction}

\begin{figure}[b]
  \centering \includegraphics{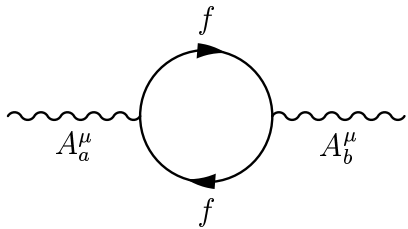}
  \parbox{140mm}{\caption{Kinetic Mixing in field theory. $f$ is a fermion carrying charge under both
      $U(1)$'s.}}
  \label{fig:km_ft}
\end{figure}
Kinetic Mixing occurs in theories that have, in addition to some visible $U(1)_a$, an additional $U(1)_b$
factor in the hidden sector. The effect occurs when the hidden $U(1)_b$ couples to the visible $U(1)_a$
through the diagram in figure \ref{fig:km_ft}. This diagram, proportional to $\mop{Tr}(Q_a Q_b)$, results
in a Lagrangian of the form
\begin{equation}
  \mathcal{L}_{\text{gauge}} = - \frac{1}{4g_a^2} F_a^{\mu\nu} F_{a\mu\nu}
                               - \frac{1}{4g_b^2} F_b^{\mu\nu} F_{b\mu\nu}
                               - \frac{\chi}{2{g_ag_b}} F_a^{\mu\nu} F_{b\mu\nu}.
\end{equation}
The consequences of this type of mixing were first studied by Holdom in the context of millicharged
particles~\cite{Holdom:1986ag}. Later, Dienes \emph{et al} pointed out that Kinetic Mixing can contribute
significantly and even dominantly to supersymmetry-breaking mediation~\cite{Dienes:1997zr} resulting in
additional contributions to the scalar mass-squared terms proportional to their hypercharge.  In this paper
we will be considering both the generation of millicharged particles (that is, particles carrying
fractionally shifted units of electric charge) and the mediation of supersymmetry breaking, in models
involving stacks of D-branes and (anti) $\overline{\text{D}}$-branes.  This is a particularly interesting
context in which to consider Kinetic Mixing because the stacks of branes and anti-branes carry $U(N)$ gauge
factors, so that Kinetic Mixing naturally occurs between these groups.

\begin{figure}[t]
  \centering \includegraphics{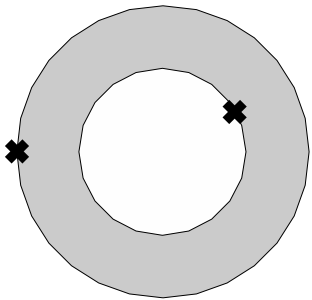}
  \parbox{140mm}{\caption{Kinetic Mixing in string theory: Annulus diagram with two open-string vertex
      operator insertions.}}
  \label{fig:annulus}
\end{figure}

The string equivalent of the Kinetic Mixing diagram is shown in figure \ref{fig:annulus}. This is a
non-planar annulus diagram, with the states in the loop corresponding to open strings stretched between the
``MSSM branes'' and anti-branes across the bulk.  However, going to the closed string channel, it can also
be seen as a closed string tree-level diagram with closed string dilaton, graviton and RR fields
propagating in the bulk. This is simply gravitational mediation and as such one expects the Kinetic Mixing
parameter $\chi$ to receive the same suppression as occurs for the other effects that have already been
discussed in the literature.  Therefore, before evaluating the diagram in detail (which we do in the
following section), let us first use dimensional arguments to estimate the expected relative strengths of
various effects.

Effects that are propagated through closed string modes in the bulk suffer a suppression of order $Y^{p-7}$
where $Y$ is the interbrane separation in units of the fundamental scale. Thus we expect
\begin{equation}
  \label{eq:chi_answer}
  \frac{\chi}{g_ag_b} \sim \mop{Tr}(\lambda_a)\mop{Tr}(\lambda_b) V_\text{NN} Y^{p-7},
\end{equation}
where $V_\text{NN}$ is the ($p-3$ dimensional) world-volume of the $p$ branes in the compactified space.
The prefactors are the traces of the Chan-Paton matrices, and the Kinetic Mixing is therefore between the
central $U(1)$'s of the $U(N)$ gauge groups. These vanish only if the gauge group on either the branes or
anti-branes is orthogonal. Note that there is no dependence on the string coupling in this expression as it
is a one loop (open string) diagram.  Now consider a set-up where the interbrane separation and
compactification scales are all of the same order, $R$, in fundamental units. The compactification scale
and string scale are related by the Planck mass which can be obtained by dimensional reduction of the 10D
theory; specifically~\cite{Ibanez:1998rf},
\begin{equation}
  \label{eq:IMR0}
 \frac{M_s}{M_P} \sim \alpha_p R^{(p-6)},
\end{equation}
where $\alpha_p$ is the fine-structure constant on the brane. The latter can be set to be of order one (it
is after all supposed to correspond to some Standard Model value) by adjusting the string coupling to
compensate for the potentially large $V_\text{NN}$ factor. We then have
\begin{equation} 
  \label{eq:chi_IMR_answer}
  \frac{\chi}{g_ag_b} \sim \left( \frac{M_s}{M_P} \right)^{\frac{2(5-p)}{6-p}}.
\end{equation}
Experimental upper bounds on $\chi$ are presented in~\cite{Davidson:2000hf}. Assuming that the hidden
sector contains some massless fields, the relevant bound is ${\chi} \simm{<} 2 \times 10^{-14}$. Inserting
into the above we find that we require $M_s\simm{<} 10^8\,\text{GeV}$ for $p=3,4$, while for $p\geq 5$ it
is impossible to avoid the millicharged particle bounds.

The above limit holds if the hidden $U(1)$ symmetry remains unbroken.  If the hidden $U(1)$ is broken then
one expects a different kind of effect to be important, namely a supersymmetry breaking $D$-term VEV of
order $M_s^2$ that can be communicated to the visible sector via the Kinetic Mixing terms. It is easy to
see that such terms would generally dominate in communicating supersymmetry breaking, as they do in the
heterotic case~\cite{Dienes:1997zr}. The potential due to the brane-antibrane attraction goes as $Y^{p-7}$
and so the corresponding effective supersymmetry breaking mass-squareds go as $\partial^2 V/\partial Y^2
\sim Y^{p-9}$.  By contrast the SUSY breaking mass-squareds communicated by Kinetic Mixing go simply as
$Y^{p-7}$, and so are dominant. The expected SUSY breaking terms in the visible sector are then of order
\begin{equation}
  \label{eq:susy_km0}
  m_\text{KM}^2 \sim M_s^2 \frac{\chi}{g_ag_b} \sim  M_s^2\left(\frac{M_s}{M_P}\right)^{\frac{2(5-p)}{6-p}}.
\end{equation}
Requiring that $ m_\text{KM}^2 \simm{<} M_W^2$ gives a similar bound on the string scale, $M_s\simm{<}
10^8\,\text{GeV}$ for $p=3,4$.  When the bound is saturated, susy breaking terms of order $M_W$ are induced
by $\overline{D3}$-branes or $\overline{D4}$-branes in the bulk.

One class of models to which our analysis is particularly relevant are the so-called intermediate scale
models~\cite{Benakli:1998pw, Burgess:1998px, Leontaris:1999qq, Antoniadis:1999xk, Aldazabal:2000sk,
  Aldazabal:2000sa, Abel:2000bj, Angelantonj:2000xf, Rabadan:2000ma, Bailin:2000kd, Bailin:2001ia,
  Cerdeno:2001se, Allanach:2001qe, Burgess:2001vr, Yang:2001ub, Alday:2002uc, Abel:2002az,
  Kiritsis:2003mc}. In these models the string scale $M_s$ is assumed to be of order $M_s \sim \sqrt{M_W
  M_P} \sim 10^{11} \,\text{GeV}$. One adopts a brane configuration that \emph{locally} reproduces the
spectrum of the MSSM but which breaks supersymmetry globally by for example the inclusion of
$\overline{\mathrm{D}}$-branes somewhere in the bulk of the compactified space. (Such supersymmetry
breaking configurations may still be consistent with the constraints of RR tadpole cancellation.) In this
set-up, the large Planck mass is a result of the dilution of gravity by a large bulk volume as usual. As
above, supersymmetry breaking communication is realised as interactions between
$\overline{\mathrm{D}}$-branes in the bulk and visible sector MSSM branes. This communication gets the same
volume suppression that gives the four-dimensional $M_P$, and so purely dimensional arguments have led to
the conclusion that supersymmetry breaking terms of order $M_{\cancel{\text{SUSY}}} \sim M_s(M_s/M_P) \sim
M_W$ are induced in the visible sector. Indeed, this effect corresponds precisely to the $Y^{p-9}$
suppression above.  However Kinetic Mixing terms can drastically modify this picture. If they are present a
more natural fundamental scale would be $M_s\sim 10^8\,\text{GeV}$.

It may seem odd that in the end an {\em upper} bound on the string scale is obtained. To see why, note that
there are two competing effects. The first obvious effect is that high string scales generate larger
supersymmetry breaking. However, the overwhelming effect for the bounds is that low string scales require
larger compactification volumes to generate the correct effective Planck mass. Consequently low string
scales allow a greater brane-antibrane separation and a reduced Kinetic Mixing.

In the following sections we give a more detailed exposition of these bounds, beginning with a study of
Kinetic Mixing between D$p$-branes and {$\overline{\mathrm{D}p}$}-branes in a simplified type II set-up.
One particular aspect that needs some attention is the question of NS-NS tadpoles which are generally
present in nonsupersymmetric set-ups. We will also discuss what happens in configurations that have
asymmetric compactification radii. This is important for two reasons. Firstly, because making some
directions transverse to the brane much smaller than the overall brane separation modifies the rate of
fall-off.  However, we find that when we recalculate the above constraints the degenerate case we have
outlined above is the optimum one, in the sense that the bounds obtained are the least restrictive.
Secondly, one would like to avoid having a too large string coupling since then the perturbative
calculation (i.e. based on strings propagating in a D-brane background) breaks down. Because of this the
compact world-volumes of the branes ($V_\text{NN}$) typically need to be much smaller than the large
transverse volumes required to dilute gravity, since the former also dilute the gauge couplings. For
D$3$-branes our conclusions are the same as above (since $V_\text{NN}=1$) and the degenerate case is
optimal. If $p>3$, the conclusions can be somewhat different: a large string scale may be recovered for
extremely asymmetric dimensions ($R/r \simm{>} 10^9$ and larger).  However, this loophole is not
particularly helpful since most models constructed to date contain $\overline{\text{D}3}$-branes, where the
more restrictive bounds apply.  Hence our main conclusions will indeed be as presented above.


\section{String calculation of Kinetic Mixing}

\subsection{Generalities}
\label{sec:cylinder}

In this section we carry out a calculation of Kinetic Mixing in a simplified type II set-up. Many features
of the end result have to do with volume dilution and are common to all brane/antibrane exchanges taking
place in a compact space.  Much of the important behaviour can therefore already be seen in the ``partition
function'', a factor in our final answer. In particular by looking at the partition function we can see how
Kaluza-Klein modes and winding modes reproduce the volume dilution one intuitively expects in both
degenerate and asymmetric compactifications. We also discuss the NS-NS tadpole which is uncancelled and
which in a perfect world would be treated by modifying the background. (This is not a perfect world.)
Finally we include the vertex factors. The calculation is done mostly in the open string channel; some of
the technicalities regarding Green functions on an annulus are presented in appendices.

Consider a setup consisting of parallel $\mathrm{D}p$ and $\overline{\mathrm{D}p}$ branes a distance $Y$
apart. Let each of these branes have an open string stuck to it, representing a $U(1)$ gauge boson. The two
open strings interact by exchanging a closed string cylinder, which we map to an annulus with two vertex
operators inserted on the boundary (fig.  \ref{fig:annulus}).  Let coordinates on the worldsheet be defined
by $z = \sigma^1 + \mi \sigma^2$, with $\sigma^1 \in [0,\pi]$ worldsheet space and $\sigma^2 \in [0,2\pi
t]$ Euclideanised worldsheet time. From the spacetime point of view, $t \rightarrow 0$ corresponds to a
long cylinder, and $t \rightarrow \infty$ a long strip. A formal expression for the amplitude is
\begin{equation}
  \label{eq:s_allint}
  \Pi^{\mu\nu}(k_1,k_2) = \sum_{\stackrel{\text{\scriptsize spin}}{\text{structures}}}
                          \int_0^\infty \frac{\dif t}{2t}
                          \left\langle  b^1(0) c^1(0) \int \! \dif z_1 \, \dif z_2 \;
                          \mathcal{V}^\mu(k_1,z_1) \mathcal{V}^\nu(k_2,z_2) \right\rangle_{\mathrm{C_2}}
\end{equation}
with $b^1(0)$ and $c^1(0)$ the spatial components of ghost fields, and $\mathcal{V}^\mu(k,z)$ vertex
operators. The factor of $2t$ corrects for the discrete symmetry coming from interchanging the ends of the
cylinder plus the continuous translational symmetry around the annulus.

We need to perform the path integral. In theory, this can be done directly. However, calculation of the
partition function
\begin{equation}
  \label{eq:z_pi}
  Z_{p\overline{p}} = \sum_{\stackrel{\text{\scriptsize spin}}{\text{structures}}}
                      \int \frac{\dif t}{2t} \left\langle b^1(0) c^1(0) \right\rangle_{\mathrm{C_2}}
\end{equation}
is easier in an operator formalism. The full amplitude can then be obtained by inserting appropriate
contributions from the vertex operators.

In a non-compact spacetime, $Z_{p\bar{p}}$ can be obtained by taking the familiar result for the amplitude
between two $p$-branes separated by a distance $Y$ in their co-volume~\cite{Polchinski:1998rq,
  Polchinski:1998rr}, and accounting for the relative flip of RR charge between a brane and antibrane by
flipping the sign of the term coming from RR closed strings.  Measuring distances in terms of the string
length (i.e. setting $l_s = \sqrt{2\alpha'}=1$),
\begin{align}
  \label{eq:z_ppbar}
  Z_{p\overline{p}} &= \frac{\mi V_4}{(2\pi)^4} \int_0^\infty \frac{\dif t}{4t}
    t^{-\frac{1}{2}(p+1)} \me^{-tY^2/\pi} \eta(\mi t)^{-12} \left[
    \underbrace{\vartheta_{00}(0|\mi t)^4 - \vartheta_{10}(0|\mi t)^4}_{\text{NS-NS strings}} +
    \underbrace{\vartheta_{01}(0|\mi t)^4}_{\text{RR strings}} \right]
\end{align}
Unlike the result for parallel $p$-branes this does not vanish, reflecting the fact that the
brane-antibrane configuration breaks all spacetime supersymmetries. Using the expansions given in appendix
\ref{sec:theta}, it can be shown that the small-$t$ limit of this result gives an attractive potential
between the branes that goes as $1/Y^{7-p}$~\cite{Polchinski:1995mt}. On the other hand, the large-$t$
limit gives a divergence from the tachyon at $Y < \pi$ which is associated with annihilation of the
brane-antibrane system at small $Y$~\cite{Banks:1995ch}.


\subsection{The partition function in compact spacetimes}

We introduce the Kinetic Mixing calculation with a prototypical set-up in which spacetime has the topology
$\mathcal{M}_4 \times T^2 \times T^2 \times T^2$. We require the branes to fill $\mathcal{M}_4$ and allow
them separations $Y_i$ in the six compact dimensions.  Modifications must be made to (\ref{eq:z_ppbar}) to
account for the compact nature of some of the dimensions. In particular, for a noncompact dimension,
integration over the string zero modes contributes a factor $V/2\pi t^{1/2}$ (where $V$ is formally
infinite). For a noncompact dimension of size $2\pi R$, we have a sum over Kaluza-Klein modes,
\begin{align}
  \label{eq:kk_zerosum}
  \sum_{n=-\infty}^{\infty}\me^{-\pi t n^2/R^2}
  = \begin{cases}
     1 \quad&:\quad t/R^2 \rightarrow \infty\\  
     R/ t^{1/2} \quad&:\quad t/R^2 \rightarrow 0
    \end{cases}
\end{align}
The second limit is obtained from the Poisson resummation formula $\sum_{n=-\infty}^{\infty}\me^{-\pi a n^2
  + 2\pi \mi b n} = a^{-1/2} \sum_{m=-\infty}^{\infty}\me^{-\pi(m-b)^2/a}$.

Strings occupying dimensions where the boundary conditions on the brane are Dirichlet can also wind around
that dimension, if it is compact. This stretching is just an extension of the brane separation term in
(\ref{eq:z_ppbar}):
\begin{equation}
  \label{eq:winding_sum}
  \me^{-tY^2/\pi}
  \rightarrow \sum_{w=-\infty}^{\infty}\me^{-t(Y+2\pi R w)^2/\pi}
\end{equation}
In each dimension, the winding modes can also be Poisson resummed,
\begin{equation}
  \label{eq:winding_resum}
  \sum_{w=-\infty}^{\infty}\me^{-t(Y+2\pi R w)^2/\pi}
    = \frac{1}{Rt^{1/2}} \left[1 + \sum_{m>0} 2\cos\left(m \frac{Y}{R}\right)
                                   \me^{-\pi m^2/4 R^2 t}\right]
\end{equation}
In the limit $R^2t \rightarrow 0$, we can take only the leading term.

As in the noncompact case, we can examine the amplitude in the large and small-$t$ limits. First, let us
examine the large-$t$ limit (which we denote as $t > t_c \simeq \pi$, since $\me^{-t/\pi}$ is our expansion
parameter).  Kaluza-Klein modes do not contribute at large $t$, so after expanding the $\vartheta$ and
$\eta$ functions, we have
\begin{equation}
  \label{eq:z_compact_larget}
  Z_{p\overline{p}} \simeq \frac{\mi V_4}{(2\pi)^4} \int_{t_c}^\infty \sdif{t} t^{-3}
    \sum_{w \in \Gamma_{9-p}} \me^{-t\left((Y+2\pi R w)^2-\pi^2\right)/\pi}
\end{equation}
where $w$ is a $9-p$ vector of integers that sums over the integer lattice $\Gamma_{9-p}$, and we are also
treating $Y$ and $R$ as vectors. Since we are here interested in $Y \gg \pi$ we can neglect the $w \in
\Gamma_{9-p} - \{0\}$ contribution and retain only the zeroth mode, giving
\begin{equation}
  \label{eq:z_compact_larget_int}
  Z_{p\overline{p}} \simeq \frac{\mi V_4}{(2\pi)^4} t_c^{-2}
                           E_3\left(t_c(Y^2-\pi^2)/\pi\right)
\end{equation}
where $E_n(z) = \int_1^\infty \sdif{t} t^{-n} \me^{-zt}$ is the standard exponential integral function.
Note that this function diverges when its argument is negative, so we still get the usual appearance of a
tachyon when $Y < \pi$.

The remainder of the partition function is evaluated in the small $t$ limit $t < t_c$, and this is where we
will find interesting results. The Kaluza-Klein modes do now contribute, and so after expanding
(\ref{eq:z_ppbar}) we have
\begin{equation}
  \label{eq:z_compact_smallt}
  Z_{p\overline{p}} = \frac{\mi V_4 V_\text{NN}}{(2\pi)^4}
                      \int_0^{t_c} \sdif{t} t^{-\frac{1}{2}(p-5)}
                      \sum_{w \in \Gamma_{9-p}} \me^{-t(Y+2\pi Rw)^2/\pi}
\end{equation}
We have written $V_\text{NN}=\left(\prod_{i=4}^p R_i\right)$ for the volume of the compact space occupied
by the branes. There is a tadpole term coming from a closed string infrared divergence (i.e. the limit
$t\rightarrow 0$).  In this limit, we can deal with the sum over windings by Poisson resumming all
dimensions, giving
\begin{equation}
  \label{eq:z_compact_smallt_resummed}
  Z_{p\overline{p}} = \frac{\mi V_4}{(2\pi)^4} \frac{V_\text{NN}}{V_\text{DD}}
                      \int_0^{t_c} \sdif{t} t^{-2}
                      \prod_{i=p+1}^9 \frac{1}{2}
                      \left[1 + \sum_{m_i>0} 2\cos\left(m_i \frac{Y_i}{R_i}\right)
                                 \me^{-\pi m_i^2/4 R_i^2 t}\right]
\end{equation}
where $V_\text{DD}=\left(\prod_{i=p+1}^9 R_i\right)$ is the volume of the compact space transverse to the
branes. Cutting off $t > 1/\mu^2$ and taking $t_c \rightarrow \infty$ gives a tadpole divergence of order
$\mu^2$ from the leading term:
\begin{equation}
  \label{eq:z_compact_smallt_cutoff}
  Z_\text{tadpole} = \frac{\mi V_4}{(2\pi)^4}
                     \frac{V_\text{NN}}{2^{9-p}V_\text{DD}}                     
                     \mu^2
\end{equation}
This expression famously corresponds to the propagation of a massless closed string state and we will
discuss it shortly.  Before we do so, let us deal with the remaining, ``threshold'', contribution where we
are on a surer footing.  To address these we need to be more discerning when deciding to Poisson resum a
particular dimension in eq.(\ref{eq:z_compact_smallt}).  First, note that for $p<5$ the integrand is
dominated by peaks at
\begin{equation}
  \label{eq:peaks}
  t_s = \frac{(5-p)\pi}{2(Y+2\pi Rw)^2}
\end{equation}
\begin{figure}[t]
  \centering
  \includegraphics{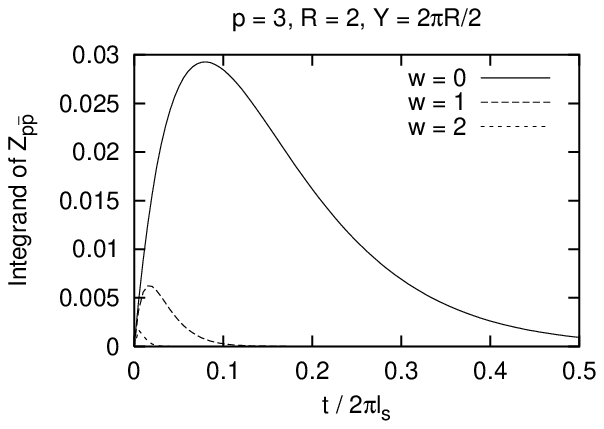}
  \hspace{1cm}
  \includegraphics{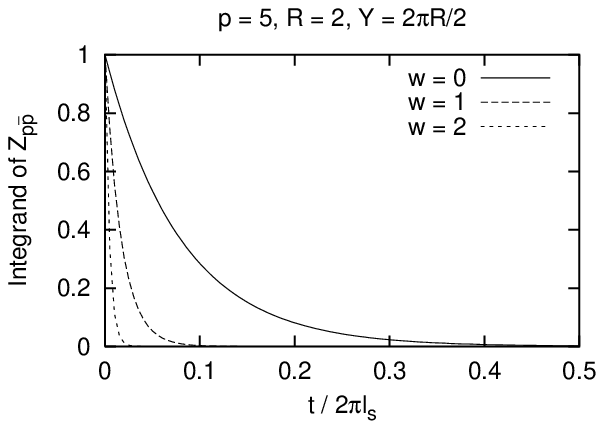}
  \parbox{140mm}{\caption{Relative contribution of different winding modes to $Z_{p\overline{p}}$, for
  $p=3$ and $p=5$. Even if $Y = 2\pi R/2$, we see that the zero-winding
  contribution is strongly dominant.}\label{fig:peaks}}
\end{figure}
Figure \ref{fig:peaks} illustrates this. The magnitude of the peaks is exponentially suppressed as the
winding number increases, and so the threshold contribution is dominated by the first peak, corresponding
to strings stretched a distance $Y$ with no winding. The tadpole divergence we saw earlier comes from an
infinite number of these peaks piling up on the origin.

\begin{figure}[t]
  \centering \includegraphics{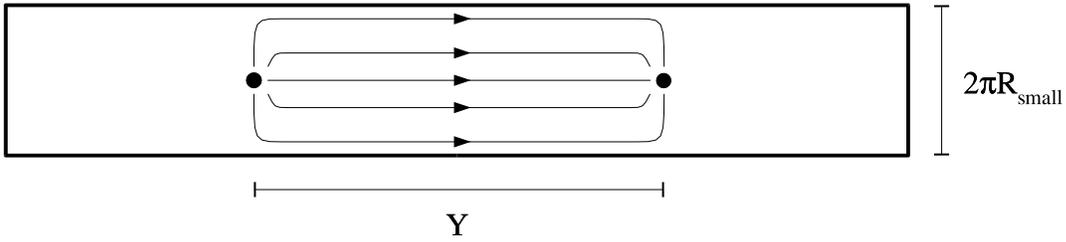} \parbox{140mm}{\caption{When to resum winding modes; a
      dimension that has $R_i \ll |Y|$ ceases to contribute to the exchange of closed string modes between
      the branes as lines of flux are confined. Winding modes in that dimension are not
      resummed.\label{fig:fluxinabox}}}
\end{figure}

Next we note that, from (\ref{eq:winding_resum}), resummation in a given dimension $i$ is only valid when
$R_i^2 t \ll 1$. Since we know that the dominant value of $t$ will be the $t_s$ given above, we should
resum the windings in a given dimension only if
\begin{equation}
  \label{eq:when_to_resum}
  R_i \ll |Y|
\end{equation}
The familiar physical situation corresponding to resumming or not is shown in figure \ref{fig:fluxinabox}. 

\vspace{3ex}
\noindent\underline{\em Degenerate radii}\\

\noindent If we take all radii to be degenerate, $R_i \equiv R$, we see that successful resummation
requires $R < Y$, which cannot be true. Hence, we choose instead to impose a small cut-off $w_0$ on the
winding lattice, indicating that we take just the leading terms.

Also, note that the larger $Y$ the further $t_s$ is below $t_c$ and the better the approximation of small
$t$ asymptotics. We can therefore neglect the large-$t$ contribution (\ref{eq:z_compact_larget}) (which is
an exponentially small correction) in the $Y \gg \pi$ limit, and take $t_c \rightarrow \infty$ in
(\ref{eq:z_compact_smallt}). The threshold corrections then look like
\begin{align}
  \label{eq:z_thr}
  Z_\text{threshold} &\approx \frac{\mi V_4 V_\text{NN}}{(2\pi)^4}
                      \int_0^\infty \sdif{t} t^{-\frac{1}{2}(p-5)}
                      \sum_{|w| < |w_0|} \me^{-t(Y+2\pi Rw)^2/\pi}\notag\\
                     &=\frac{\mi V_4 V_\text{NN}}{(2\pi)^4} 
                      \Gamma\left(\tfrac{1}{2}(7-p)\right) \sum_{|w|<|w_0|}
                      \left(\frac{\pi}{(Y+2\pi Rw)^2}\right)^{\frac{1}{2}\left(7-p\right)}
\end{align}

For $p \geq 5$ both the divergence and the threshold terms are maximal at $t=0$ so it is less obvious that
this prescription is correct for them also. To see that it is correct, it is useful to consider the tadpole
contribution in the un-resummed ``many winding'' picture. Consider the $|w| \gg 1$ contribution to
$Z_\text{threshold}$ that we have removed, with $Y=0$ for convenience:
\begin{align}
  \label{eq:z_inf}
  Z_\infty &=\frac{\mi V_4 V_\text{NN}}{(2\pi)^4} 
             \Gamma\left(\tfrac{1}{2}(7-p)\right) \sum_{|w|>|w_0|}
             \left(\frac{\pi}{(2\pi Rw)^2}\right)^{\frac{1}{2}\left(7-p\right)}
\end{align}
We may approximate the divergent sum involved by an integral, imposing a large cutoff $w_\mu$ on the
winding,
\begin{align}
  \label{eq:infsum}
  \sum_{|w_0|<|w|<|w_\mu|} \frac{1}{|2\pi Rw|^{(7-p)}}
    &\approx \int_{w_0}^{w_\mu} \frac{\Omega_{8-p} w^{8-p} \dif{w}}{(2\pi Rw)^{(7-p)}}\notag\\
    &= \frac{\Omega_{8-p}}{(2\pi R)^{9-p}} \frac{1}{2}\left[(2\pi R w_\mu)^2 - (2\pi R w_0)^2\right]
\end{align}
where $\Omega_{8-p}$ is the area of a unit $(8-p)$-sphere. Then,
\begin{align}
  \label{eq:z_inf_2}
  Z_\infty &=\frac{\mi V_4}{(2\pi)^4} \frac{V_\text{NN}}{2^{9-p} V_\text{DD}}
             \Gamma\left(\tfrac{1}{2}(7-p)\right) \pi^{-\tfrac{1}{2}(7-p)}\Omega_{8-p}
             \frac{1}{2}\left[(2R w_\mu)^2 - \left(2R w_0\right)^2\right]
\end{align}
Clearly this is the $\mu^2$ divergence of the massless closed string with
\begin{equation}
  \label{eq:mu_equiv}
  \mu^2 = \frac{1}{2} \Gamma\left(\tfrac{1}{2}(7-p)\right) \pi^{-\tfrac{1}{2}(7-p)}\Omega_{8-p} (2R w_\mu)^2
\end{equation}
The tadpole contribution may be excised by removing the contributions with many windings, leaving the
threshold contribution which corresponds to just the leading term. This picture does not rely on the
presence of the saddle.


\vspace{3ex}
\noindent \underline{\em Asymmetric radii}\\

\noindent Suppose now that the $R_i$ are not all equal, and that we have $d$ dimensions which meet the
criterion (\ref{eq:when_to_resum}) for resummation, leaving $9-p-d$ dimensions which require a cutoff on
$w$. By (\ref{eq:winding_resum}), the resummed dimensions contribute a factor $1/R_i t^{1/2}$ to
$Z_{p\bar{p}}$, so that
\begin{equation}
  \label{eq:z_thr_asymm}
  Z_\text{threshold} = \frac{\mi V_4}{(2\pi)^4}\frac{V_\text{NN}}{2^{d}V_\text{DD,small}} 
                       \Gamma\left(\tfrac{1}{2}(7-p-d)\right) \sum_{|w|<|w_0|}
                       \left(\frac{\pi}{(Y+2\pi Rw)^2}\right)^{\frac{1}{2}\left(7-p-d\right)}
\end{equation}
where $V_\text{DD,small}$ is the volume of the small Dirichlet-Dirichlet dimensions that have been
resummed. Physically, the reduction in the power to which $Y$ is raised comes from the volume restriction
on closed string modes exchanged between the branes (figure \ref{fig:fluxinabox}). Note that we need $p+d <
7$ to avoid a divergence in $Z_{p\bar{p}}$. Hence, scenarios with $d=4$ appear untenable.

We now return to the tadpole contribution. As we have seen this piece corresponds to the propagation of a
massless closed string mode. External propagators should be removed from the effective potential and the
corresponding field (i.e. the dilaton) inserted in their place. The end result is that this contribution
simply corresponds to the term
\begin{equation}
  \label{eq:effpot}
  -2 T_9 \int d^{10}x \sqrt{-G}\me^{-\Phi}
\end{equation}
in the effective potential. Thus at this stage it is not appropriate to include the tadpole contribution to
$\chi$ since in principle the contribution would be one particle reducible, corresponding to the exchange
of an on-shell massless closed string state between the two gauge fields. A more correct way to deal with
it would be a generalisation of the Fischler-Susskind mechanism~\cite{Fischler:1986ci, Fischler:1986tb}, in
which the background is modified to take account of the extra term (\ref{eq:effpot}) in the effective
potential, upon which as shown in ref.~\cite{Dudas:2000ff} "spontaneous compactification" can occur.
Intuitively one expects the propagation of the massless modes to be screened by the Fischler-Susskind
mechanism, so that the cut-off we have imposed by hand on the effective potential, $\mu \sim R w_\mu$, may
take on a physical meaning as some kind of screening length. Since in this article we do not wish to
address the Fischler-Susskind mechanism in any detail, we will simply work in the toroidal background and
use the OPI argument to excise the tadpole from $\chi$.  One should bear in mind however that volume
factors involving the radius along the $Y$ direction should possibly be understood as the size of some
spontaneously compactified dimension.


\subsection{Inclusion of vertex operators}
\label{sec:vo}

Let us now proceed to evaluate the Kinetic Mixing amplitude (\ref{eq:s_allint}) by including vertex
operators in $Z_{p\bar{p}}$. For the annulus, it is necessary that the vertex operators' superghost charges
sum to zero. Hence, we work with vertex operators in the $0$-picture, in which a $U(1)$ gauge boson
corresponds to
\begin{equation}
  \label{eq:vo_gb}
  \mathcal{V}^\mu(k,z) = \mi g_O \lambda \left(\dot{X}^\mu(z)
                       + \mi k\cdot \psi(z) \psi^\mu(z)\right)\me^{\mi k\cdot X(z)}
\end{equation}
Here, $g_O$ is the open string coupling and $\lambda$ the Chan-Paton matrix for the brane or stack of
branes associated with the vertex operator. The index $\mu$ runs over the noncompact dimensions. We have
again set $l_s=\sqrt{2\alpha'}=1$.

The correlation between two such vertex operators can be calculated either by summing over all contractions
or via a path integral approach. After integration by parts, an expression for the amplitude is
\begin{multline}
  \label{eq:amp}
  \Pi^{\mu\nu} = -g_O^2 \mop{Tr}(\lambda_a)\mop{Tr}(\lambda_b)
  \left[ k^\mu k^\nu - \eta^{\mu\nu} k^2\right] \sum_{(\alpha\beta)=(00)}^{(11)} Z_{\alpha\beta}\\
  \times \int_0^{2\pi t} \!\!\dif \sigma^2_1 \int_0^{2\pi t} \sdif{\sigma^2_2} \me^{k^2
    G(z_1-z_2)} \left[ (2\pi)^2 \frac{\partial G(z_1 - z_2)}{\partial \sigma^2_1} \frac{\partial G(z_1 -
      z_2)}{\partial \sigma^2_2} -  S_{\alpha\beta}(z_1 - z_2)^2 \right]
\end{multline}
$G(z)$ and $S_{\alpha\beta}(z)$ are the respective propagators for bosons and fermions on opposite side of
the annulus, described in Appendix \ref{sec:green}. $Z_{\alpha\beta}$ is the $(\alpha\beta)$ spin structure
term in $Z_{p\bar{p}}$. It is to be understood that the integration over $t$ contained in $Z_{p\bar{p}}$ is
applied to the entire expression. The factor of $(2\pi)^2$ comes in because distances around the annulus
are measured in multiples of $2\pi$.

Now, since the vertex operators are always on opposite edges of the annulus, we can always write $z_1 - z_2
= \pi + \mi x$, where $x$ is the difference in worldsheet time between positions of the propagators. Since
the annulus is periodic in the time direction, we can reduce the integral over both vertex operator
positions to an integral over just one, multiplied by $2\pi t$. Furthermore, it is not necessary to
integrate this remaining vertex operator over the entire annulus; by symmetry, we can just integrate
halfway round and multiply by two. With these simplifications in place, we can rewrite the second line of
(\ref{eq:amp}) as
\begin{align}
  \label{eq:cbit_details}
  -4\pi t \int_0^{\pi t} \sdif{x} \me^{k^2 G(x)}
  \left[ (2\pi)^2\partial_xG(x)^2+(2\alpha')^2S_{\alpha\beta}(x)^2 \right]
\end{align}
From experience of the cylinder diagram, we expect the interesting physics to come from small values of
$t$. Hence, we modular-transform the propagators so that their expansions will be in $1/t$:
\begin{equation}
  \label{eq:voigrand_opp}
  (2\pi)^2\partial_xG(x)^2+(2\alpha')^2S_{\alpha\beta}(x)^2 = (4\pi\alpha')^2
  \left[\frac{\partial_x^2\vartheta_4\left(\frac{x}{2\pi t}\big|\frac{\mi}{t}\right)}
             {\vartheta_4\left(\frac{x}{2\pi t}\big|\frac{\mi}{t}\right)}
      - \frac{\partial_x^2\vartheta_{ba}\left(\frac{x}{2\pi t}\big|\frac{\mi}{t}\right)\big|_{x=0}}
             {\vartheta_{ba}\left(\frac{x}{2\pi t}\big|\frac{\mi}{t}\right)\big|_{x=0}}
  \right]
\end{equation}
Note that terms involving first derivatives of the theta functions have cancelled between the bosonic and
fermionic propagators. For the analogous calculation on the torus, the spin-structure independent terms
cancel entirely~\cite{Kaplunovsky:1988rp,Kaplunovsky:1992vs,Kiritsis:1997hj}, but this does not occur on
the annulus.

We are now in a position to write down the final amplitude. As discussed in the previous section, this will
contain both a `tadpole' contribution and a `threshold' contribution. Taking the tadpole contribution
(which we will ultimately ignore) first, we have (after expanding the theta functions and integrating over
$x$),
\begin{equation}
  \label{eq:amp_tad}
  \Pi^{\mu\nu}_\text{tadpole} = 2\mi V_4 \frac{V_\text{NN}}{2^{9-p}V_\text{DD}}
                                g_O^2 \mop{Tr}(\lambda_a)\mop{Tr}(\lambda_b)
                                \left[ k^\mu k^\nu - \eta^{\mu\nu} k^2\right]
                                \int_{1/\mu^2}^\infty\sdif{t} t^{k^2-2} \me^{-\pi k^2/4t}
\end{equation}
Performing the integral and expanding in $k$,
\begin{align}
  \label{eq:amp_tad_igrl}
  \int_{1/\mu^2}^\infty\sdif{t} t^{k^2-2} \me^{-\pi k^2/4t} = \mu^2 + \mathcal{O}(k^2)
\end{align}
Taking $k^2 \rightarrow 0$ (since our gauge bosons are massless), the final result is
\begin{align}
  \label{eq:amp_tad_final}
  \Pi^{\mu\nu}_\text{tadpole} = 2\mi V_4
                                \frac{V_\text{NN}}{2^{9-p}V_\text{DD}}
                                g_O^2 \mop{Tr}(\lambda_a)\mop{Tr}(\lambda_b)
                                \left[ k^\mu k^\nu - \eta^{\mu\nu} k^2\right]
                                \mu^2
\end{align}
Let us now examine the threshold contribution. For notational simplicity, we take the degenerate radii
result. This time, we obtain
\begin{align}
  \label{eq:amp_thr_opp}
  \Pi^{\mu\nu}_\text{threshold} = &2\mi V_4 V_\text{NN}
                                  g_O^2 \mop{Tr}(\lambda_a)\mop{Tr}(\lambda_b)
                                  \left[ k^\mu k^\nu - \eta^{\mu\nu} k^2\right]\notag\\
                                  &\times \int_0^\infty\sdif{t}
                                    t^{k^2-\frac{1}{2}(p-5)} \me^{-\pi k^2/4t}
                                  \sum_{|w|<|w_0|}\me^{-t(Y+2\pi Rw)^2/\pi}
\end{align}
As before, we have imposed a small cut-off $|w_0|$ on the winding lattice to avoid the tadpole. Performing
the integral and expanding the Bessel function as in appendix \ref{sec:theta} before expanding again in
$k$, we obtain
\begin{align}
  \label{eq:amp_thr_igrl}
  \int_0^\infty\sdif{t} t^{k^2-\frac{1}{2}(p-5)} \me^{-\pi k^2/4t}\me^{-tY^2/\pi}
  &= 2\left(\frac{k\pi}{2Y}\right)^{\tfrac{1}{2}(7-p)+k^2}K_{\tfrac{1}{2}(p-7)-k^2}(kY)\notag\\
  &= \left(\frac{\pi}{Y^2}\right)^{\tfrac{1}{2}(7-p)}\Gamma\left(\tfrac{1}{2}(7-p)\right) +
  \mathcal{O}(k^2)\qquad(p<7)
\end{align}
Taking $k^2\rightarrow 0$ and ignoring all winding modes except the zeroth,
\begin{equation}
  \label{eq:amp_thr_final}
  \Pi^{\mu\nu}_\text{threshold}
               = 2\mi V_4 V_\text{NN}
                 g_O^2 \mop{Tr}(\lambda_a)\mop{Tr}(\lambda_b) \left[ k^\mu k^\nu - \eta^{\mu\nu} k^2\right]
                 \left(\frac{\pi}{Y^2}\right)^{\tfrac{1}{2}(7-p)}
                 \Gamma\left(\tfrac{1}{2}(7-p)\right)
\end{equation}
As for the partition function, the necessary modifications for asymmetric radii are to introduce a factor
$2^{d}V_\text{DD,small}$ on the bottom and to replace $p \rightarrow p + d$. Rewriting the $k\mu$ in terms
of $F^{\mu\nu}$ (with an appropriate normalisation to remove factors of the string coupling $g_O$), the
relevant parameter measuring the mixing is
\begin{equation}
  \label{eq:chi_km}
  \frac{\chi}{g_a g_b} = \frac{2}{N} \mop{Tr}(\lambda_a)\mop{Tr}(\lambda_b) \frac{V_\text{NN}}{2^{d}V_\text{DD,small}}
                         \left(\frac{\pi l_s^2}{Y^2}\right)^{\tfrac{1}{2}(7-p-d)}
                         \Gamma\left(\tfrac{1}{2}(7-p-d)\right)
\end{equation}
We have explicitly restored $l_s$ in this expression, except for the volume factors which are expressed in 
units of string lengths. 

There are two further subtleties we should mention: orbifolds, and branes of differing dimensionality.
Firstly, suppose we make our internal space $(T^2 \times T^2 \times T^2)/\mathbb{Z}_N$, and fix the branes
at two different orbifold singularities.  The resulting theory contains an untwisted sector, in which the
states are just those in the non-orbifolded theory, plus $N-1$ twisted sectors consisting of states that
survive the orbifold projection. The boundary conditions on twisted states prevent them from having
momentum~\cite{Hamidi:1987vh}, which keeps them stuck at fixed points. Hence, figure \ref{fig:annulus}
cannot occur for twisted states in the theory.  We therefore neglect twisted-sector contributions to the
amplitude, and simply divide $\Pi^{\mu\nu}$ by a factor $N$ to get the result for Kinetic Mixing with a
$\mathbb{Z}_N$ orbifold.

Secondly, suppose that instead of a D$p$-$\overline{\mathrm{D}p}$ combination, we have a setup consisting
of D$p$ and $\overline{\mathrm{D}q}$ branes (or equivalently, D$q$ and $\overline{\mathrm{D}p}$ branes),
where $q>p$. Let dimensions $i=4,\ldots,p$ be shared Neumann-Neumann, $i=p+1,\ldots,q$ be Neumann-Dirichlet
and $i=q+1,\ldots,9$ be Dirichlet-Dirichlet. The nonzero number of ND dimensions will not affect our
results since, as explained in appendix \ref{sec:green}, there is no correlation between vertex operators
in ND dimensions. The only changes to the amplitudes calculated is that $V_\text{NN}$ should be taken as
the volume of compact Neumann-Neumann dimensions \emph{shared} by the branes, and $V_\text{DD}$ should be
appropriately reduced.


\section{Millicharged particles from Kinetic Mixing}
\label{sec:milli}

We first assume that $U(1)_b$ is unbroken, so that millicharged particles are generated by Kinetic Mixing.
As we argued in section \ref{sec:cylinder}, only the amplitude contribution (\ref{eq:amp_thr_final}) is
relevant here.  One can now look at the consequences of this mixing in different scenarios, using
experimental data on the maximum size of millicharged particles. To do this, we will need the following
relation, obtained from dimensional reduction of the type I string action~\cite{Ibanez:1998rf},
\begin{equation}
  \label{eq:IMR}
  \left(\frac{M_s}{M_P}\right)^2 = 2\alpha_p^2 \frac{V_\text{NN}}{2^{9-p}V_\text{DD}}
\end{equation}
$M_s=1/l_s$ is the string scale, and $\alpha_p$ is the coupling on the brane. Since the type I theory can
be considered as an orientifold of type IIB, and type IIA is related to type IIB by T-duality, this result
ought to be valid in all brane-based models.


\vspace{3ex}
\noindent \underline{\em Degenerate radii}\\

\noindent Let us first consider the case of degenerate extra dimensions, $R_i \equiv R$ with $d=0$.  Take
the brane separation to be $Y=\pi R$ and write $g_a g_b = 4\pi\alpha_p$, so
\begin{equation}
  \label{eq:chi_km_degen}
  \chi_\text{degen} = \frac{2}{N} \mop{Tr}(\lambda_a)\mop{Tr}(\lambda_b)
         \left(\frac{2^{\tfrac{1}{2}(8-p)}}{\alpha_p}\frac{M_s}{M_P}\right)^\frac{2(5-p)}{(6-p)}
         4\pi\alpha_p \pi^{\tfrac{1}{2}(p-7)} \Gamma\left(\tfrac{1}{2}(7-p)\right)
\end{equation}
The mixing parameter $\chi$ is related to the observable charge shift $\epsilon$ simply by $\chi = -
\frac{e_a}{e_b}\epsilon$~\cite{Holdom:1986ag}. Experimental upper bounds on $\epsilon$ are examined in
\cite{Davidson:2000hf}, where it is found that for particles of mass $m_\epsilon \simm{<} m_e$, $|\epsilon|
\simm{<} 2 \times 10^{-14}$ is excluded.

We can use this information to put an upper bound on the string scale. For $p=3$, suppose we assume
$e_a/e_b \sim 1$, $\mop{Tr}(\lambda_a)\mop{Tr}(\lambda_b) \sim 1$, and take $N=3$, $\alpha_p \approx 1/24$
(the MSSM unification value).  The requirement $|\epsilon| \simm{<} 2 \times 10^{-14}$ then gives
\begin{equation}
  \label{eq:limit_p3_degen}
  M_s \simm{<} 5 \times 10^7 \, \text{GeV}
\end{equation}
By the argument at the end of section \ref{sec:vo}, the same result holds between for
D$3$-$\overline{\mathrm{D}q}$ system, where $q>3$; the only difference is that we ought to take $g_a g_b =
4\pi\sqrt{\alpha_p\alpha_q}$.

Note that the existence of millicharged particles at some level is necessary to avoid a nonzero $M_s$. If
millicharged particles do not exist at all in nature, then the only resolution is to insist either that no
unbroken $U(1)$'s exist on the antibrane, or that $\mop{Tr}(\lambda_a)\mop{Tr}(\lambda_b)$ is fortuitously
zero. This could be the case if all antibranes present have orthogonal gauge groups on their world volumes
(e.g. if they are located at orientifold planes).

Setting $p>3$ leads to similar conclusions; when $p=4$ we find $M_s \simm{<} 4 \times 10^4\,\text{GeV}$,
$p=5$ is clearly ruled out as $\chi$ has no dependence on any mass scale, $p=6$ implies $M_P > M_s$, whilst
$\chi$ becomes singular for $p\ge 7$.


\vspace{3ex}
\noindent \underline{\em Asymmetric radii}\\

\noindent Suppose we set the number of small Dirichlet-Dirichlet dimensions to be $d>0$, with the small
dimensions having radius $r$ whilst the others have radius $R$. Again set the brane separation to $Y=\pi
R$. We cannot now eliminate both radii from (\ref{eq:chi_km}) using (\ref{eq:IMR}), and choose to leave the
free parameter as the ratio $R/r$.

There are two cases to consider. First, we suppose that the dimensions wrapped by the brane are of size
$R$, i.e. $V_\text{NN}=R^{p-3}$. The result is that $\chi$ is enhanced by a power of $R/r$ relative to the
degenerate case:
\begin{equation}
  \label{eq:chi_km_asymm_1}
  \chi_\text{asymm} = \chi_\text{degen} \left(\frac{R}{r}\right)^{\frac{d}{6-p}}
    \left(\frac{\sqrt{\pi}}{2}\right)^d
    \frac{\Gamma\left(\tfrac{1}{2}(7-p-d)\right)}{\Gamma\left(\tfrac{1}{2}(7-p)\right)}
\end{equation}
Since $R>r$, this enhancement factor is always greater than one for $p<6$. Hence, we see that the
degenerate case is optimal; the bound on $M_s$ becomes more restrictive as $R/r$ increases. The conclusion
that unbroken $U(1)$'s cannot exist on the antibrane then appears unavoidable.

Alternatively, we may take $V_\text{NN} = r^{p-3}$, so that the extra dimensions wrapped by the brane are
small. This assumption is perhaps more natural given that we want to end up with gauge couplings $\sim 1$
without having an overly large string coupling. In this case, we find
\begin{equation}
  \label{eq:chi_km_asymm_2}
  \chi_\text{asymm} = \chi_\text{degen} \left(\frac{R}{r}\right)^{\frac{d-(p-3)}{6-p}}
    \left(\frac{\sqrt{\pi}}{2}\right)^d
    \frac{\Gamma\left(\tfrac{1}{2}(7-p-d)\right)}{\Gamma\left(\tfrac{1}{2}(7-p)\right)}
\end{equation}
The difference with the first case is that we have $d \rightarrow d-(p-3)$ in the exponent of $R/r$. This
shows that the small NN directions are working against the small DD directions, which is what one would
expect by T-duality. For $p=3$, things are of course the same as before since $V_\text{NN}=1$. For $p>3$,
it appears possible to suppress the Kinetic Mixing effect by having $d<p-3$. However, $R/r$ must be very
large to recover a large string scale. For $p=4$ and $d=0$, we need $R/r \sim 10^7$ to give $M_s \simm{<}
10^{11}$, for example. For $p=5$ with $d=0$ or $d=1$, $\chi$ contains no dependence on $M_P$ or $M_s$ and
so experimental data serves only to constrain $R/r \simm{>} 10^6$ or $R/r \simm{>} 10^{12}$ respectively.
Even if one accepts these large values of $R/r$, millicharged particles must be predicted at some level.


\section{SUSY breaking communication}
\label{sec:susy}

It is known that, if supersymmetry is a feature of nature, then its breaking is highly restricted if
experimental constraints are to be satisfied. The explanation for this that we most frequently cite is that
supersymmetry breaking occurs at some high scale in a hidden sector and is communicated to the visible
sector by some process which both weakens it and gives it the desired form. Intermediate-scale brane models
contain hidden antibranes which are present to ensure cancellation of Ramond-Ramond tadpoles.  These
provide natural candidates for the hidden sector.

Reference~\cite{Dienes:1997zr} showed, in the context the heterotic string, that if we suppose some physics
causes a hidden $U(1)_b$ to break, then Kinetic Mixing is a candidate for the mediation process. The result
is an additional contribution to supersymmetry breaking mass-squareds in the visible sector of the form
\begin{equation}
  \label{eq:km_susybreaking}
  m_\text{KM}^2 = g_a Q_a \chi \langle D_b \rangle
\end{equation}
Identifying $U(1)_a=U(1)_Y$, this results in extra supersymmetry breaking terms proportional to
hypercharge. The authors of~\cite{Dienes:1997zr} also pointed out that it cannot be the only source of
mediation, as some of the mass-squareds would have to be negative, and these authors therefore focused on
placing an upper limit on $\chi$ in order to avoid destabilising the gauge hierarchy (\emph{i.e.} to avoid
supersymmetry breaking in the visible sector much larger than 1 TeV).  The appropriate limit on $\chi$ then
depends on the scale of supersymmetry breaking in the hidden sector which in turn depends on the other
sources of mediation (\emph{e.g.} gravity or gauge). The conclusion was that generic models with gravity
mediation would have disastrously large Kinetic Mixing if the hidden sector contained additional $U(1)$'s.
The relevant bound to avoid destabilising the hierarchy is $\chi \simm{<} 10^{-16}$. Such a small coupling
constitutes a fine tuning according to the criterion of t'Hooft.  In heterotic strings the situation can be
ameliorated somewhat because the gauge groups are usually unified into some non-abelian GUT groups. The
Kinetic Mixing only arises due to mass splittings once the GUT groups are broken, and one finds typical
values of $\chi \sim 10^{-9}$; much less than 1 but still large enough to destabilize the hierarchy.

Let us apply the same considerations to non-supersymmetric D-brane configurations.  In intermediate scale
models, supersymmetry is usually assumed to be broken by annulus diagrams with no vertex operators: this is
supersymmetry breaking in the bulk. In this case, $Z_{p\bar{p}}$ should be treated as a potential felt by
observers on the visible brane due to the presence of the antibrane. The term in an effective Lagrangian
with dimensions of mass squared is then $M_s^2 (\partial^2{Z_{p\bar{p}}}/\partial{Y^2})$, and so
supersymmetry breaking terms are of the order
\begin{equation}
  \label{eq:susy_cylinder}
  m_{\cancel{\text{SUSY}}}^2 \sim M_s^2 \frac{\partial^2 Z_{p\bar{p}}}{\partial Y^2}
                             \sim M_s^2 \frac{V_\text{NN}}{Y^{9-p}}
                             \sim M_s^2 \frac{V_\text{NN}}{V_\text{DD}}
                             \sim \frac{M_s^4}{M_P^2}
\end{equation}
We have used (\ref{eq:IMR}) in the last step, and ignored extraneous factors.

However if supersymmetry is broken on the antibrane it will be communicated across to the visible sector by
Kinetic Mixing. Let us now suppose that there is a $U(1)_b$ present on the antibrane, and that some physics
causes the $D$-term of this $U(1)$ to acquire a VEV.  The scale of supersymmetry breaking is then
\begin{equation}
  \label{eq:susy_km}
  m_\text{KM}^2 \sim M_s^2 \chi
\end{equation}
where $\chi$ includes just the threshold contributions. For asymmetric dimensions with
$V_\text{NN}=R^{p-3}$, we find
\begin{equation}
  \label{eq:susy_km_2}
  m_\text{KM}^2 \sim M_s^2 \left(\frac{M_s}{M_P}\right)^\frac{2(5-p)}{(6-p)}
                           \left(\frac{R}{r}\right)^{\frac{d}{6-p}}
\end{equation}
The key point is that the usual bulk breaking contribution goes as $1/M_P$, whereas the Kinetic Mixing
contribution receives less suppression. Of the two effects then, Kinetic Mixing will always be dominant if
it is present (i.e. if $\mop{Tr}(\lambda_a)\mop{Tr}(\lambda_b) \ne 0$). In that case, requiring
$m_\text{KM} \sim M_W$ gives us
\begin{equation}
  \label{eq:susy_limit}
  M_s \sim M_P \left(\frac{M_W}{M_P}\right)^{\frac{6-p}{11-2p}}
               \left(\frac{r}{R}\right)^{\frac{d}{2(11-2p)}}
\end{equation}
We are led to the conclusion that the string scale must be much lower than the usual $M_s \sim 10^{11}
\,\text{GeV}$ in order to generate the right sort of visible supersymmetry breaking in the visible sector.
For instance, with degenerate extra dimensions and $p=3$, we find $M_s \sim 10^8 \, \text{GeV}$. Including
all numerical factors from (\ref{eq:chi_km}, \ref{eq:km_susybreaking}) raises this by perhaps an order of
magnitude, but the general conclusion is the same. In most cases, asymmetric extra dimensions only serve to
lower the string scale further. The exception is if the visible sector is a $p>3$-brane when one might try
to circumvent this restriction by identifying $V_\text{NN} = r^{p-3}$ and demanding a large value of $R/r$,
as discussed above in the context of millicharged particles. In this case, one obtains
(\ref{eq:susy_limit}), but with $d \rightarrow d-(p-3)$. However, $R/r$ must be extremely large to recover
$M_s \sim 10^{11} \,\text{GeV}$: for $p=4$ we need $R/r \simm{>} 10^{20}$, and for $p=5$, $R/r \simm{>}
10^9$.

One might alternatively assume that some unrelated effect such as gaugino condensation is responsible for
supersymmetry breaking, and that Kinetic Mixing must be a sub-dominant contribution in order to avoid
destabilising the hierarchy. In that case $M_\text{KM} < M_W$ is required, and the $\sim$ in equation
(\ref{eq:susy_limit}) becomes a $\simm{<}$, giving an upper bound on the string scale.


\section{Summary and conclusions}

Kinetic Mixing provides an opportunity to constrain non-supersymmetric D-brane configurations (e.g.
intermediate scale models) using current phenomenological data. We have shown that the effect will
generically occur between the visible branes and hidden antibranes present in such models, and that it will
have observable consequences for low-energy physics. It can be avoided only if all antibranes present have
orthogonal gauge groups on their world volumes (e.g. if they are located at orientifold planes).  This is a
stringent demand on the global configuration.

From experimental limits on millicharged particles, we have shown that in intermediate-scale brane models,
one must generally either accept a string scale which is $M_s \simm{<} 10^{8} \,\text{GeV}$, or require
that there be no unbroken $U(1)$'s on the antibrane. If we accept a lower string scale or take advantage of
the discussion at the end of section \ref{sec:milli}, we then predict millicharged particles at some level.
The consequences with broken $U(1)$'s in the hidden sector are similar. If we are to avoid destabilising
the hierarchy, we must either accept $M_s \simm{<} 10^{8} \, \text{GeV}$, or again ensure
$\mop{Tr}(\lambda_a)\mop{Tr}(\lambda_b)$ always vanishes.

The overall conclusion must be that intermediate-scale models -- and indeed, any model containing branes
and antibranes -- are more strongly constrained than was previously thought. It is interesting that a
strong upper bound on the string scale is obtained, as this pushes the models in a direction where they are
likely to conflict with other constraints due to excessively large instanton or Kaluza-Klein couplings
(e.g.  refs.~\cite{Cvetic:2003ch, Abel:2003vv, Abel:2003fk, Abel:2003yx, Ghilencea:2002da}).  This upper
bound is a result of the large volumes required to dilute the effect of Kinetic Mixing.

Construction of phenomenologically realistic models consistent with the demands of Kinetic Mixing remain an
interesting avenue for investigation. Apart from configurations that have antibranes with only orthogonal
groups, one possibility which is quite attractive is to set the string scale at $\sim 10^8\,\text{GeV}$ and
to use the Kinetic Mixing mediation to generate mass-squared terms in the visible sector that are
proportional to hypercharge. This results in a significant amelioration of the so-called flavour problem of
supersymmetry in the effective $N=1$ model. The problem to be addressed here however would be how to
prevent negative mass-squareds. One alternative, to have a second non-anomalous visible sector $U(1)$, does
not seem to arise very naturally in the models that have been constructed to date, but may be worth
investigating.


{\bf\centering Acknowledgements\\}\vspace{2ex}

It is a great pleasure to thank Sacha Davidson, Emilian Dudas and John March-Russell for several helpful
discussions. This research was supported by a PPARC studentship, and by Opportunity Grant
PPA/TS/1998/00833.


\appendix

\section{Some special function properties}
\label{sec:theta}

The $\vartheta$ and $\eta$ functions possess a defined behaviour under modular transformations:
\begin{align}
  \label{eq:theta_modular}
  \vartheta_{\alpha\beta}\left(\frac{\mi x}{2\pi},\mi t\right)
  &=\mi^{\alpha\beta}t^{-1/2}\me^{x^2/4\pi t}\vartheta_{ba}\left(\frac{x}{2\pi t},\frac{\mi}{t}\right)\\
  \eta(\mi t) &= t^{-1/2}\eta(\mi/t)
\end{align}
This allows us to expand them in terms of $t$ or in terms of $1/t$. Here we just give the expansion in
terms of $1/t$,
\begin{align}
  \label{eq:theta_expansions_1_over_t}
  \vartheta_{11}\left(\frac{x}{2\pi t},\frac{\mi}{t}\right)
   &= 2\me^{-\pi/4t}\sin(x/2t) - 2\me^{-9\pi/4t}\sin(3x/2t) + 2\me^{-25\pi/4t}\sin(5x/2t) + \cdots\\
  \vartheta_{10}\left(\frac{x}{2\pi t},\frac{\mi}{t}\right)
   &= 2\me^{-\pi/4t}\cos(x/2t) + 2\me^{-9\pi/4t}\cos(3x/2t) + 2\me^{-25\pi/4t}\cos(5x/2t) + \cdots\\
  \vartheta_{00}\left(\frac{x}{2\pi t},\frac{\mi}{t}\right)
   &= 1 + 2\me^{-\pi/t}\cos(x/t) + 2\me^{-4\pi/t}\cos(2x/t) + 2\me^{-9\pi/t}\cos(3x/t) + \cdots\\
  \vartheta_{01}\left(\frac{x}{2\pi t},\frac{\mi}{t}\right)
   &= 1 - 2\me^{-\pi/t}\cos(x/t) + 2\me^{-4\pi/t}\cos(2x/t) - 2\me^{-9\pi/t}\cos(3x/t) + \cdots\\
   \eta(\mi/t)^{-m} &= \me^{m\pi/12t} + m\me^{(m-24)\pi/12t} + \cdots
\end{align}
The Bessel K function appears in (\ref{eq:amp_thr_opp}). For non-integral $\nu$, this has the small-$z$
expansion
\begin{align}
  \label{eq:besselk}
  K_{\nu}(z) &= \frac{1}{2}\left(\frac{z}{2}\right)^{-\nu}
                \left[\Gamma(\nu) - \left(\frac{z}{2}\right)^2 \Gamma(\nu-1) + \cdots \right]\notag\\
             &+ \frac{1}{2}\left(\frac{z}{2}\right)^{\nu}
                \left[\Gamma(-\nu) - \left(\frac{z}{2}\right)^2 \Gamma(-\nu-1) + \cdots \right]
\end{align}
As $z\rightarrow 0$, the first set of terms end up dominating for negative $\nu$ and the second for
positive $\nu$.


\section{Green functions on the annulus}
\label{sec:green}

\subsection{Bosonic}

\begin{figure}[tb]
  \centering
  \includegraphics{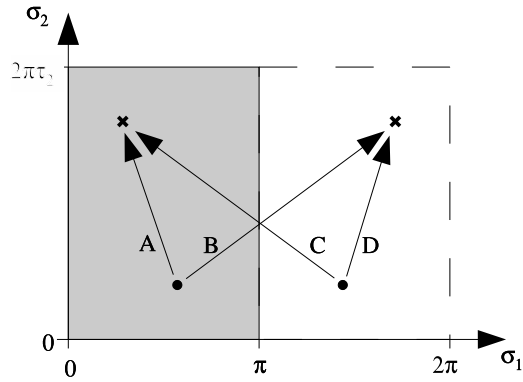}
  \caption{The annulus as obtained from the torus by setting $\tau_1=1$ and identifying under
    $\sigma^1=-\sigma^1$ (i.e. $z=-\bar{z}$). A fundamental region is shown shaded. The propagator
    is obtained by supplementing each transition (marked A,D) by an image charge piece (marked B,C).}
  \label{fig:torusannulus}
\end{figure}

A torus may be defined as a region of the complex plane $ds^2 = \dif z \dif \bar{z}$ with periodic boundary
conditions
\begin{equation}
  \label{eq:torus_pbe}
  z = z + 2\pi(m + n\tau) \qquad\qquad m,n \in \mathbb{Z}
\end{equation}
where $z = \sigma^1 + \mi \sigma^2$ and $\tau = \tau_1 + \mi\tau_2$. This describes a parallelogram in the
complex plane. If we set $\tau_1=1$ so that the parallelogram becomes a square, then we have only one
modulus $t \equiv \tau_2$. Then, identifying $z = - \bar{z}$ gives us the annulus (fig.
\ref{fig:torusannulus}).  Effectively, what we have done is to score around two circles in the plane of the
torus, and fold the result back on itself.

Using this identification, we can obtain the propagator on an annulus from that on a torus. On a torus
parameterised as described, the propagator is\footnote{The reader who is familiar
  with~\cite{Polchinski:1998rq} should note that the propagator given there is slightly different, as it
  does not include a contribution from the self-energy of the fields (see e.g.~\cite{Kostelecky:1988px,
    Headrick:Chapter7}).}~\cite{Burgess:1987ah}
\begin{align}
  \label{eq:g_torus}
   G^{T^2}(z)
    &= - \frac{\alpha'}{2} \ln \left| \frac{\vartheta_1 \left( \frac{z}{2\pi} \big| \tau \right)}
                                           {\eta(\tau)^3} \right|^2
       + \alpha' \frac{[\mathrm{Im}(z)]^2}{4\pi \tau_2}
\end{align}
Here, $\vartheta_1(\nu | \tau)$ and $\eta(\tau)$ are the Jacobi theta- and Dedekind eta-functions, as
defined in~\cite{Polchinski:1998rq}.

We have here ignored an irrelevant additive contribution related to the compact nature of the space, since
this always drops out of the final amplitude. Now, by including image charges as shown in figure
\ref{fig:torusannulus}, we can get the propagator on an annulus.

In general, we have a choice of Neumann or Dirichlet conditions on the worldsheet boundaries. About a
Neumann boundary, the open string mode expansion is symmetric, whilst about a Dirichlet boundary it is
antisymmetric. We can impose N or D boundary conditions on our propagator by requiring it to have similar
characteristics. In our situation, open string vertex operators are confined to the two branes and their
have momenta parallel to both.  Hence, NN boundary conditions are appropriate; we require our propagator be
symmetric under both $z_1 \rightarrow -\bar{z_1}$ and $z_2 \rightarrow -\bar{z_2}$.
\begin{equation}
  \label{eq:g_ann_nn}
  G(z_1-z_2) = \left[ G^{T^2}(z_1-z_2|\mi t) + G^{T^2}(z_1+\bar{z_2}|\mi t) \right]
\end{equation}
For vertex operators on opposite sides of the annulus, we take $z_1 = 0 + \mi \sigma^2_1$, $z_2 = \pi + \mi
\sigma^2_2$. Defining $x \equiv \sigma^2_1 - \sigma^2_2$ and using $\vartheta_1(\nu \pm \frac{1}{2} | \tau)
= \vartheta_2(\nu | \tau)$, plus the fact that the appropriate theta-functions are always real for purely
imaginary $\nu$, $\tau$,
\begin{equation}
  \label{eq:g_specific}
  G(x) = - 2 \alpha' \ln \left[ \frac{\vartheta_2 \left( \frac{\mi x}{2\pi} \big| \mi t \right)}{\eta(\mi
      t)^3} \right] + \alpha' \frac{x^2}{2\pi t}
\end{equation}


\subsection{Fermionic}

The fermion propagator on a torus with periodic boundary conditions can be obtained by noting that the
derivative of a solution to the bosonic action solves the Dirac action. For the three `even' spin
structures, where the boundary conditions are not periodic, the torus is artificially extended by a factor
of two in both directions and the method of images is used~\cite{Burgess:1987wt}. The result is
\cite{Kiritsis:1997hj}
\begin{equation}
  \label{eq:s_evenspin}
  S_{\alpha\beta}^{T^2}(z) = 2\pi \frac{\vartheta_{\alpha\beta}(\frac{z}{2\pi},\tau)\vartheta_1'(0|\tau)}
                        {\vartheta_1(\frac{z}{2\pi}|\tau)\vartheta_{\alpha\beta}(0,\tau)}
\end{equation}
We can convert these to the annulus by the method of images as before. There is an extra complication due
to the way in which the involutions applied to the torus exchange left- and right-moving fermions, which
requires us to insert gamma-matrices appropriately~\cite{Burgess:1987wt}. For Neumann-Neumann boundary
conditions on the annulus,
\begin{align}
  \label{eq:s_ann_nn}
  \mathbf{S}(z_1-z_2) &= \frac{1}{2}\Big[ \mathbf{S}(z_1-z_2) + \gamma_1\mathbf{S}(-\bar{z_1}-z_2)
    + \mathbf{S}(z_1+\bar{z_2})\gamma_1^T + \gamma_1\mathbf{S}(-\bar{z_1}+\bar{z_2})\gamma_1^T
                                    \Big] \notag\\
  &= \begin{pmatrix} S^{T^2}(z_1-z_2) & S^{T^2}(z_1+\bar{z_2}) \\
                     S^{T^2}(z_1+\bar{z_2}) & S^{T^2}(z_1-z_2) \end{pmatrix}
\end{align}
In the last line we have used the result that $S(z)=S(-\bar{z})$ for our torus propagator, which comes
directly from the theta-function identity $\vartheta_{\alpha\beta}(\nu,\tau) =
\vartheta_{\alpha\beta}(-\bar{\nu},\tau)$. Attempts to construct a propagator with either
Dirichlet-Dirichlet or Neumann-Dirichlet boundary conditions fail, as all terms then cancel.

The off-diagonal terms represent propagators between fermions moving in opposite directions, and are not of
interest to us. Thus, we have found that on the annulus with Neumann conditions on the boundaries, the
propagator is the same as that for a torus:
\begin{equation}
  \label{eq:s_annulus_nn_2}
  S(z_1 - z_2) = S^{T^2}(z_1 - z_2)
\end{equation}
Now, using the identity~\cite{Mumford:1982}
\begin{equation}
  \label{eq:theta_identity}
  \left[\frac{\vartheta_{\alpha\beta}(z,\tau)\vartheta_1'(0|\tau)}
             {\vartheta_1(z|\tau)\vartheta_{\alpha\beta}(0,\tau)}\right]^2
    = \frac{\partial_z^2\vartheta_{\alpha\beta}\left(\frac{z}{2\pi}\big|\mi t\right)\big|_{z=0}}
           {\vartheta_{\alpha\beta}\left(\frac{z}{2\pi}\big|\mi t\right)\big|_{z=0}}
    - \frac{\partial^2}{\partial z^2}\log\vartheta_1(z,\tau)
\end{equation}
we can write for the even spin structures, with vertex operators on opposite boundaries of the annulus as
before,
\begin{equation}
  \label{eq:s_evenspin_ann}
  S_{\alpha\beta}(x)^2 = -4\pi^2
  \left[ \frac{\partial_x^2\vartheta_{\alpha\beta}\left(\frac{\mi x}{2\pi}\big|\mi t\right)\big|_{x=0}}
              {\vartheta_{\alpha\beta}\left(\frac{\mi x}{2\pi}\big|\mi t\right)\big|_{x=0}}
        - \frac{\partial^2}{\partial x^2}\log\vartheta_2\left(\frac{\mi x}{2\pi}\Big|\mi t\right) \right]
\end{equation}


\nocite{*}

\bibliography{km}
\bibliographystyle{h-elsevier2}

\end{document}